\documentclass[fleqn,twoside,fps]{article}
\usepackage{espcrc2,graphicx,latexsym,amssymb,amsmath}
\usepackage[figuresright]{rotating}

\title{Chiral symmetry breaking and topology for all $N$}

\author{N. Cundy\address[Wup]{Department of Physics, University of Wuppertal, 
        Gaussstrasse 20, D--42097 Wuppertal, Germany},
        M. Teper\address[Oxf]{Theoretical Physics, Oxford
        University, 1 Keble Road, Oxford OX1 3NP, United Kingdom},
        U. Wenger\addressmark[Oxf]\thanks{Talk
presented by U.W. at Lattice 2003.}}

\begin{document}

\begin{abstract}
We investigate spontaneous chiral symmetry breaking in SU($N$) gauge
theories at large $N$ using overlap fermions. The exact zero modes and
the low-lying modes of the Dirac operator provide the tools to gain
insight into the interplay between chiral symmetry breaking and
topology. We find that topology indeed drives chiral symmetry breaking
at $N=3$ as well as at large $N$. By comparing the results on various
volumes and at different lattice spacings we are able to show that our
conclusions are not affected by finite volume effects and also hold in
the continuum limit. We then address the question whether the topology
can be usefully described in terms of instantons.
\end{abstract}

\maketitle

\section{INTRODUCTION}
The SU(3) gauge fields of QCD possess non-trivial topological
properties and these properties are related to the zero modes of the
Dirac operator. The importance of the non-trivial topology partly
stems from the fact that it leads to the axial U(1) anomaly and
therefore to a massive $\eta'$ in the chiral limit. On the other hand,
topology is not necessarily involved in the spontaneous breaking of
chiral symmetry. Through the Banks-Casher relation we know that the
chiral condensate is proportional to the density of small eigenmodes
of the Dirac operator, $\langle \bar \psi \psi\rangle \sim
\lim_{\lambda \rightarrow 0} \rho(\lambda)$. A popular scenario is
therefore to assume that the non-vanishing density is due to exact
zero modes of the Dirac operator which interact with each other and
are lifted away from zero eigenvalue producing in this way
$\lim_{\lambda \rightarrow 0} \rho(\lambda) \neq 0$. These near-zero
modes would therefore have a topological origin since they emerged
from the topological zero modes.

It is an interesting and important question to ask whether this
scenario is indeed true for SU(3) and, moreover, whether it remains
valid in the large $N$ limit \cite{wenger}. While we believe that the
chiral symmetry of QCD is spontaneously broken also at $N=\infty$, it
is not clear at all whether the topological charge density is
localised, as suggested by semi-classical considerations, or whether
it corresponds to a 'continuous' topological charge distribution due
to large vacuum fluctuations from confinement, as argued by Witten
\cite{witten}.

In order to address these questions we resort to a non-perturbative
lattice calculation where we examine the local chiral and topological
structures of the low-lying fermion modes. The modes serve as probes
of the non-perturbative vacuum that are insensitive to ultraviolet
fluctuations and therefore they allow us to determine their
topological content.

As an important side remark we note that any statement we are able to
make in quenched QCD at large $N$ readily applies to full QCD at large
$N$ as well.

\section{LATTICE CALCULATION}
\begin{figure*}[t]
\begin{center}
\vspace{-1cm}
\begin{eqnarray*}
\includegraphics[angle=-90,width=7.25cm]{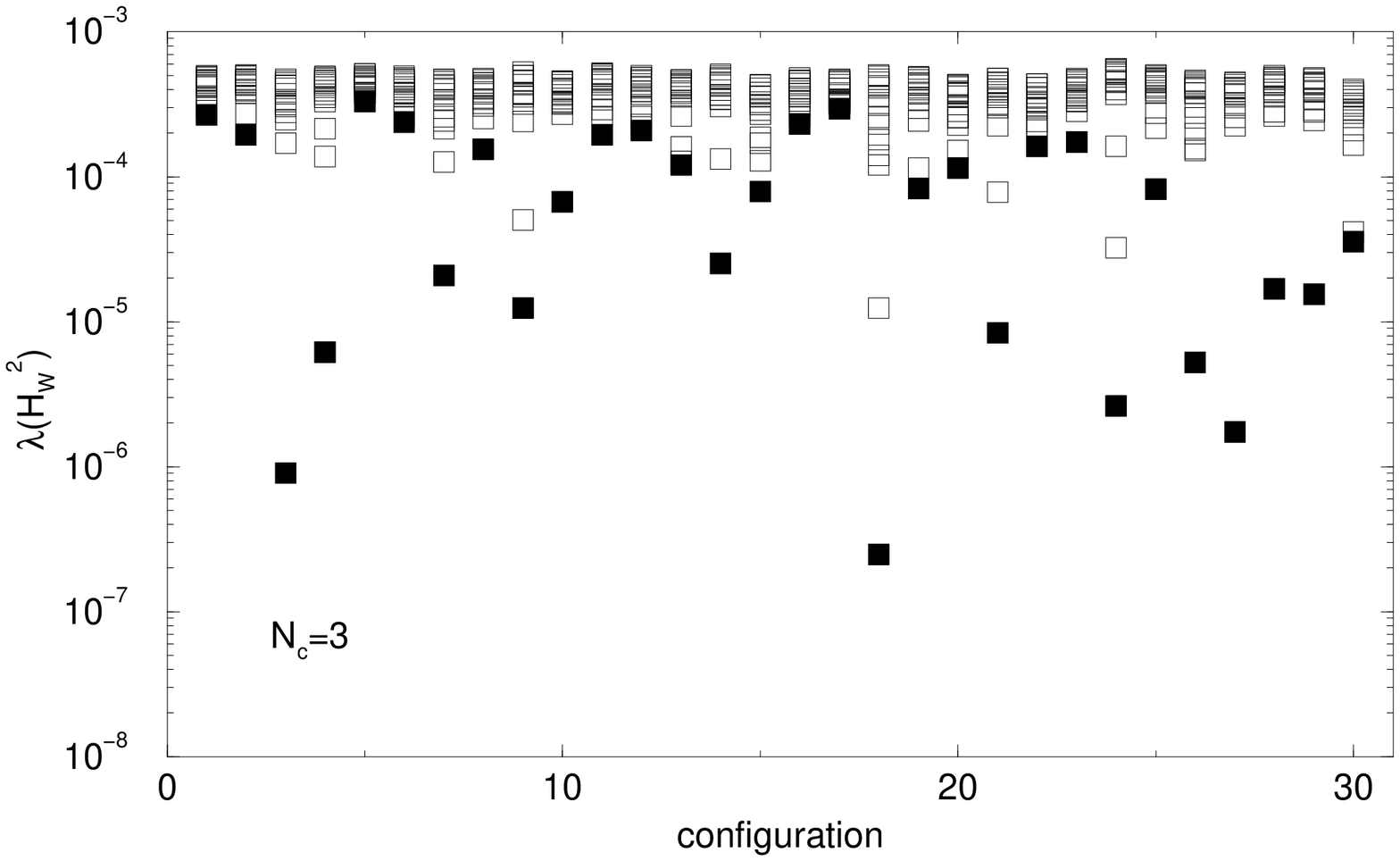} &
\includegraphics[angle=-90,width=7.25cm]{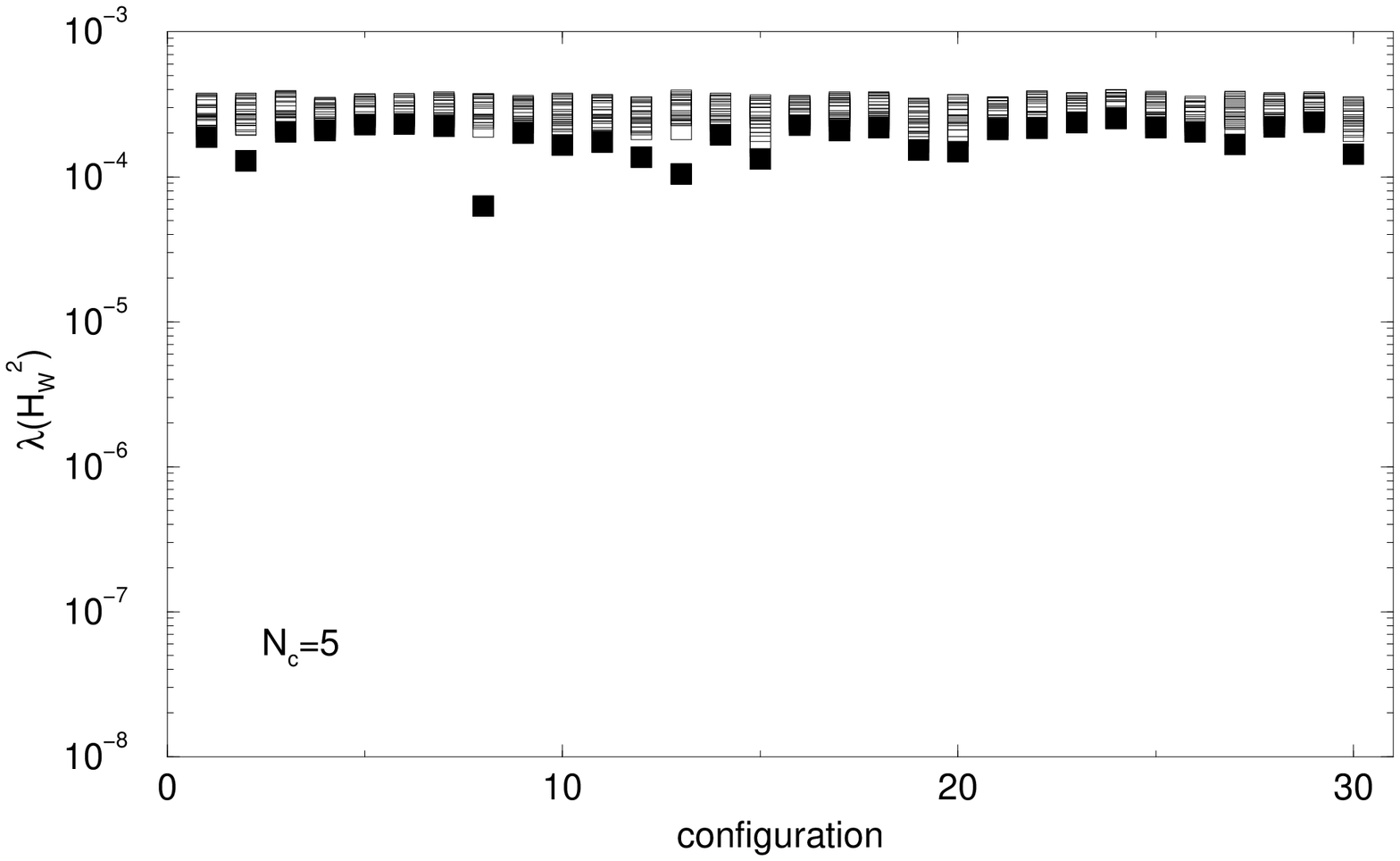}
\end{eqnarray*}
\end{center}
\vspace{-0.8cm}
\caption{{}Lowest 15 eigenvalues of $H_{\text W}^2$ at fixed lattice
spacing $a\simeq0.09$ fm and fixed volume $V=4.0 \, \text{fm}^4$ for
$N=3$ (left) and 5 (right).}
\vspace{-0.2cm}
\label{fig:Hw_eigenvalues}
\end{figure*}
We generated three ensembles of gauge field configurations using the
pure gauge Wilson action with $N=2,3,4,5$ at two different lattice
spacings $a \simeq 0.12$ and $0.09$ fm and two different volumes $V
\simeq 4.0$ and $13.6 \, \text{fm}^4$. The lattice spacing is set by
the string tension $a \sqrt{\sigma} = 0.261$ and $0.196$,
respectively, for all gauge groups.  We then calculated all eigenmodes
of the chirally symmetric overlap Dirac operator $D(0) \psi_\pm =
\lambda_\pm \psi_\pm$ with $\text{Im} \lambda_\pm < 520$ MeV.

We observe very large autocorrelation of the topological charge at
small $a$ and large $N$. It is due to the suppression of topological
charge fluctuations at the cut-off (dislocations) and was already
observed in \cite{wenger}. It is also reflected in the eigenvalue
distribution of the hermitian Wilson Dirac operator $H_{\text W}$
entering the overlap operator construction. The striking effect of the
disappearance of dislocations, and consequently the small eigenvalues,
is exemplified in fig.~\ref{fig:Hw_eigenvalues} where we compare the
lowest 15 eigenvalues $\lambda(H_{\text W}^2)$ for the ensembles with
$a=0.09$ fm and $V=4 \, \text{fm}^4$ at $N=3$ and 5.  The behaviour
indicates that for large enough $N$ the cost for the overlap operator
will follow the naively expected scaling with $V$ and $N$, i.e.,
$\propto N^2$ and $N^3$ for the fermionic and gauge part,
respectively. This is in contrast to the SU(3) case where we encounter
additional costs due to the increasing number of small eigenvalues
$\lambda(H_{\text W}^2)$. Together with the intriguing finite size
scaling in \cite{lucini}, this motivates a revisit of the large $N$
reduction using overlap fermions \cite{neuberger}.

\section{RESULTS}
The chiral condensate and thus the eigenmode density is expected to
scale with $V$ and $N$. While we observe the correct volume scaling,
the correct scaling with $N$ begins to set in only at the finer
lattice spacing. Furthermore we seem to observe large scaling
violations with $a$ for this quantity.

In order to investigate now the interplay between topology and chiral
symmetry breaking we employ the following strategy. Using some measure
we determine the topological content of the near-zero modes and use
the zero modes as a comparison since we know that these are
topological in origin. To be more specific we define the chiral
density $\omega_5(x) = \psi^\dagger(x) \gamma_5 \psi(x)$ and the
topological charge density $q(x)$ of the background gauge fields
obtained after cooling and calculate the dimensionless overlap
\cite{cundy}
\begin{multline}
	C_d^5 = \int d^4x |\omega_5(x)|^d 
	|q(x)|^{1-d} \\
	\times \text{sign}(\omega_5(x)) \text{sign}(q(x)).  
\end{multline}
The advantage of using such a definition is that it is scale invariant
for any $d$ as long as only one scale is involved in both $\omega_5$
and $q$.

For SU(3) we find that the overlap is comparable for the zero modes
and the near-zero modes and that it decreases as the eigenvalues of
the near-zero modes increase.  This remains qualitatively true as $V
\rightarrow \infty$ and $a \rightarrow 0$. We therefore conclude that
topology indeed drives chiral symmetry breaking in SU(3) and that the
influence of topology is weakening for higher lying modes.

\begin{figure}[t]
\includegraphics[angle=-90,width=7.25cm]{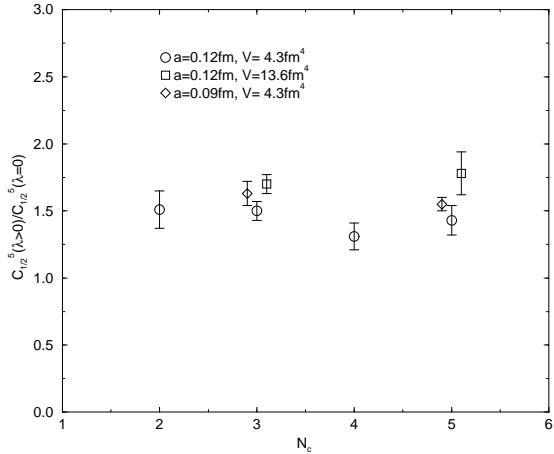}
\vspace{-0.7cm}
\caption{{}Ratio between the scale invariant overlaps of the zero and
near-zero modes for $d=1/2$.}
\vspace{-0.5cm}
\label{fig:ovlaps}
\end{figure}

As $N$ grows we find that the overlaps become smaller for both the
zero modes and the near-zero modes, and that this happens
approximately at the same rate. As a consequence the ratio between
overlaps from the zero modes and near-zero modes is roughly constant
for all $N$ (see fig.~\ref{fig:ovlaps}). Again, this remains
qualitatively true as $V \rightarrow \infty$ and $a \rightarrow
0$. All these findings suggest that topology drives chiral symmetry
breaking for all $N$, despite the fact that the local chirality
becomes weaker as $N$ grows.

\section{INSTANTONS$\ldots$?}
In order to investigate whether instantons can usefully describe
topology at large $N$, we repeated the calculations on cooled
configurations as well as on artificial semi-classical instanton
configurations. Again we calculated the lowest few eigenmodes,
determined the scale invariant overlaps, the local chiral and
topological structures and the instanton size distributions
from the chiral densities of the eigenmodes. As $N$ grows we find that
small instantons (dislocations) are suppressed and that the typical
instanton size grows. Furthermore the eigenmodes become rapidly less
chiral while the gauge fields themselves are less (anti-)selfdual. We
therefore conclude that instantons seem to become less useful for
describing topology at large $N$, despite the fact that topology
remains important.

An interesting point to note, however, is the following
\cite{schafer,shuryak}. Several instantons which overlap strongly in
coordinate space may interact only weakly when they occupy mutually
commuting SU(2) subgroups of SU($N$). So the density of instantons
grows $\propto N$ and the local chiral density of the eigenmodes as
well as the (anti-)selfduality of the gauge fields appear to become
weaker, although the instanton liquid remains dilute in the sense that
the interactions are weak. Calculations using different SU(2) subgroup
projections seem to indicate that this possibility is indeed partly
realised, however, its effect appears to be very weak (at least for
the gauge groups we have considered).

Without commenting on its relevance for SU($N$) Yang-Mills we note as
a curious side remark, that in ${\cal N}=4$ SUSY Yang-Mills at
$N=\infty$ the so-called 'master field', i.e.~the field configuration
dominating the path integral, is an instanton cluster in which all
instantons share a common location and size \cite{dorey}.

\section{CONCLUSIONS} 
We investigated the role of topology in the breaking of chiral
symmetry in SU($N$) gauge theories. Calculations for $N=2,3,4,5$ allow
us to make statements about all $N$. We calculated the low lying
eigenmodes of the Dirac operator which drive chiral symmetry breaking
and determined their topological content. We performed calculations at
two lattice spacings and two volumes so as to have some control over
the corresponding corrections. We obtain convincing evidence that
topology does indeed drive chiral symmetry breaking for SU(3) and that
this remains so for all $N$. We find that dislocations are suppressed
and that the local chirality and the (anti-)selfduality of the vacuum
becomes weaker as $N$ grows. Whether the topology can be usefully
described in terms of instantons is thus not at all clear.

\end{document}